\documentclass[aps,prd,reprint,nofootinbib,amsmath,amssymb]{revtex4-2}

\usepackage{graphicx} % Para incluir figuras (espectros QNM, potenciales)
\usepackage{hyperref} % Para hipervínculos en referencias y ecuaciones
\usepackage{color}
\usepackage{multirow}
\usepackage{booktabs} % Opcional, pero recomendado para tablas elegantes
\usepackage{amsmath}  % Para mejorar el renderizado de símbolos matemáticos
\usepackage{siunitx} % Para alineación perfecta de números
\usepackage{subcaption}

\begin{document}

\title{Quasinormal Modes of Schwarzschild-MOG Black Holes: Scalar, Electromagnetic, and Coupled Vector-Tensor Perturbations via WKB and PINNs}

\author{Daniel Andrés Rojas Ordoñez}
\email{anrojasor@unal.edu.co}
\author{Eduard Larrañaga}
\email{ealarranaga@unal.edu.co}
\affiliation{Universidad Nacional de Colombia, Sede Bogotá. Facultad de Ciencias. Observatorio Astronómico Nacional. Ciudad Universitaria. Bogotá, 111321, Colombia.}

\date{\today}

\begin{abstract}

In the era of gravitational wave astronomy, the damping phase in the evolution of perturbed black holes provides an unparalleled tool to test General Relativity  and contrast it with alternative theories. This study investigates the spectrum of quasinormal modes of black holes within the (Scalar-Tensor-Vector) MOG gravity framework, under the influence of coupled scalar, electromagnetic, and gravitational perturbations. To evaluate the complex frequencies, we implement a dual methodological approach: the semi-analytical WKB approximation extended up to the sixth order and the use of Physics-Informed Neural Networks (PINNs). By rigorously establishing the effective electro-gravitational coupling in the perturbed background, we demonstrate that the MOG deviation parameter $\alpha$ monotonically reduces both the real oscillation frequency and the decay rate across all perturbation sectors. Additionally, we show that PINNs reconstruct the global spatial structure of the fundamental eigenfunctions and reproduce the WKB fundamental frequencies with small discrepancies in most regimes, the largest deviations appearing in the imaginary part of the coupled vector-tensor modes at strong coupling and in the scalar monopole. These findings not only correct previous spectral estimates in the literature, but also provide distinct phenomenological signatures that could help constrain MOG using future gravitational wave observations.

\end{abstract}

\maketitle

% ---------------------------------------------------------
% SECCIÓN I: INTRODUCCIÓN
% ---------------------------------------------------------
\section{Introduction}
\label{sec:intro}

The advent of gravitational wave astronomy has opened a new observational window on the universe and provides a powerful means to test General Relativity (GR) in highly dynamical, strong-field conditions. Following the historic detections of binary black hole mergers by the LIGO and Virgo collaborations \cite{abbott2016gw151226, abbott2016observation, abbott2017gw170104, abbott2017gw170814}, an unprecedented astrophysical laboratory for the strong gravity regime has been consolidated. Recently, research has turned to study the possibility of detecting perturbation processes in which, for example, a massive test particle plunges into a supermassive black hole from slightly below the innermost stable circular orbit (ISCO). During the ringdown phase, the system generates a multipolar gravitational waveform composed of a superposition of a large number of quasinormal modes (QNMs), whose frequencies are, in general, complex numbers \cite{kokkotas1999quasi}. The real part of these frequencies dictates the oscillation of the mode, while the imaginary part determines its temporal decay rate. According to the No-Hair Theorem of GR, these frequencies depend exclusively on the mass and spin of the final black hole; therefore, QNMs act as a unique spectral fingerprint. Any detected deviation in this spectrum would imply the existence of additional degrees of freedom, making the \textit{ringdown} the ideal scenario to discriminate between Einstein's formulation and alternative theories of gravity \cite{abbott2016tests, berti2009quasinormal, Konoplya_2011, sibandze2017echoesblackholesevidence, Lasky_2010}. \\

The Scalar-Tensor-Vector Gravity (STVG) theory, commonly referred to as Modified Gravity (MOG), developed by J. W. Moffat \cite{moffat2006scalar, Moffat_2015, moffat2017misalignedspinmergingblack} is a covariant extension of general relativity. Its fundamental motivation is to offer a unified gravitational description that explains galactic dynamics and cosmological observations without postulating the existence of dark matter \cite{Moffat_2013, moffat2015cosmological}. To achieve this, MOG enriches the gravitational sector by introducing a massive spin-1 vector field ($\phi_{\mu}$) that couples a repulsive fifth force to baryonic matter and promotes certain fundamental constants to dynamical scalar fields. Under these premises, by solving the vacuum field equations, it is possible to derive exact solutions that generalize the Schwarzschild and Kerr metrics within the framework of MOG \cite{moffat2015black}. The Schwarzschild-MOG solution is mathematically similar to the Reissner-Nordström metric, but the crucial distinction lies in the underlying physics: the charge present in the MOG spacetime is of a strictly gravitational nature, induced by the intrinsic vector field of the theory. \\

Although the quasinormal mode spectrum of MOG black holes has been previously explored in the literature \cite{manfredi2018quasinormal}, a rigorous reassessment of the electro-gravitational coupling within the perturbation framework is required. Previous analyses have overlooked the full parametric mapping induced by the enhanced gravitational constant, leading to a decoupling between the perturbation equations and the actual modified background geometry. Addressing this analytical inconsistency is very important for obtaining a physically accurate spectral signature that can be reliably compared with future gravitational wave data.\\

In this work, a phenomenological and numerical analysis is developed on the quasinormal mode spectrum of the Schwarzschild-MOG black hole subjected to coupled scalar, electromagnetic, and gravitational perturbations. To accurately evaluate the complex frequencies and extract the continuous physical wavefunctions, a dual computational approach is implemented. On the one hand, the sixth-order WKB semi-analytical technique with Padé approximants is used to ensure the convergence of the frequency spectrum \cite{konoplya2026efficient, konoplya2019higher}. On the other hand, Physics-Informed Neural Networks (PINNs) \cite{raissi2019physics, cornell2022using, jin2020unsupervised} are applied, to our knowledge for the first time in this context. While the WKB method provides excellent perturbative estimates for the eigenvalues, the PINN framework allows for the direct resolution of the regularized differential equations, overcoming the divergent asymptotic boundaries and reconstructing the global spatial topology of the eigenfunctions.\\

The remainder of this paper is organized as follows. In Section \ref{sec:MOG_intro}, we briefly review the STVG/MOG theory and the static, spherically symmetric black hole solution. In Section \ref{sec:perturbations}, we derive the master equations governing the test scalar, electromagnetic, and coupled gravitational perturbations, establishing the corrected effective mapping for the electro-gravitational coupling. Section \ref{sec:methodology} details the numerical implementation of the 6th-order WKB method and the architecture of the Physics-Informed Neural Network. The numerical results for the complex frequencies and the reconstructed eigenfunctions are discussed in Section \ref{sec:results}. Finally, in Section \ref{sec:conclusions}, we summarize our main physical findings and address the computational scope of our methods. Throughout this work, we adopt signature (+ - - -).

\section{The Static Black Hole in Modified Gravity (MOG)}
\label{sec:MOG_intro}

The MOG theory was developed by J. W. Moffat \cite{moffat2006scalar} as a covariant extension of General Relativity. Its main motivation is to offer a unified gravitational description that explains galactic dynamics, velocity dispersion in clusters, and cosmological observations without the need to postulate the existence of dark matter \cite{Moffat_2013, Moffat_2014, Moffat_2015, moffat2015cosmological, moffat2014scalar, Shojai_2017}. To achieve this phenomenology, MOG enriches the gravitational sector by introducing new dynamical degrees of freedom coupled to the metric tensor $g_{\mu\nu}$. Specifically, it postulates the existence of a massive spin-1 vector field ($\phi_\mu$), and promotes certain fundamental constants to dynamical scalar fields: the gravitational constant $G(x)$, the mass of the vector field $\mu(x)$, and its coupling constant. The total action of the system is expressed as the sum of the gravitational, vector, scalar, and matter contributions \cite{moffat2006scalar},
\begin{equation}
    S = S_G + S_V + S_S + S_M.
\end{equation}

The standard tensor sector is modified by allowing $G$ to vary in spacetime, resulting in a generalized Einstein-Hilbert action,
\begin{equation}
    S_G = \frac{1}{16\pi} \int d^4x \sqrt{-g} \left[ \frac{1}{G} (R + 2\Lambda) \right],
\end{equation}
where $R$ is the Ricci scalar and $\Lambda$ is the cosmological constant. Meanwhile, the dynamics of the vector field is governed by a Maxwell-Proca type Lagrangian, which couples a repulsive force (fifth force) to baryonic matter,
\begin{equation}
    S_V = - \int d^4x \sqrt{-g} \left[ \frac{1}{4} B^{\mu\nu}B_{\mu\nu} - \frac{1}{2}\mu^2 \phi^\mu \phi_\mu \right],
\end{equation}
where $B_{\mu\nu} = \partial_\mu \phi_\nu - \partial_\nu \phi_\mu$. Finally, the scalar action $S_S$ contains the kinetic terms and the self-interaction potentials $V(G, \mu)$ for the fields $G(x)$ and $\mu(x)$. However, we will not consider these contributions here.\\

From the MOG field equations, it is possible to derive an exact solution that generalizes the Schwarzschild metric by considering that the gravitational coupling  $G = G_N(1+\alpha)$ is an effective constant \cite{moffat2015black} and introducing the gravitational charge $Q_g = \sqrt{\alpha G_N} M$. Additionally, the mass of the particle associated with the vector field $\phi_\mu$ ($m_\phi \sim 10^{-28}$ eV) is neglected \cite{moffat2015black, Moffat_2013}. Under these premises, the fundamental field equations are expressed as follows
\begin{align}
    R_{\mu\nu} &= -8\pi G\, T^{(\phi)}_{\mu\nu}, \label{eq:ricci_mog} \\
    \frac{1}{\sqrt{-g}}\partial_\nu(\sqrt{-g}B^{\mu\nu}) &= 0, \label{eq:maxwell_like1} \\
    \partial_\sigma B_{\mu\nu} + \partial_\mu B_{\nu\sigma} + \partial_\nu B_{\sigma\mu} &= 0, \label{eq:maxwell_like2}
\end{align}
where the generalized energy-momentum tensor for the vector field, $T^{(\phi)}_{\mu\nu}$, is defined by
\begin{equation}
    T^{(\phi)}{}_{\mu \nu} = -\frac{1}{4\pi} \left( B_{\mu\alpha}B^{\; \alpha}_\nu - \frac{1}{4}g_{\mu\nu} B^{\alpha\beta}B_{\alpha\beta} \right). \label{eq:stress_energy}
\end{equation}

By solving this coupled system, the invariant line element for a static and spherically symmetric spacetime takes the following form:
\begin{equation}
    ds^2 = f(r) dt^2 -  f(r)^{-1} dr^2 - r^2 d\Omega^2,
\label{eq:metric_mog}
\end{equation}
where 
\begin{equation}
    f(r) = \left( 1 - \frac{2G_N(1+\alpha)M}{r} + \frac{\alpha(1+\alpha)G_N^2 M^2}{r^2} \right).
\end{equation}

Notably, this metric is mathematically similar to the Reissner-Nordström solution, which describes a black hole with electric charge. The crucial distinction lies in the underlying physics: the charge $Q_g > 0$ present in equation \eqref{eq:metric_mog} is strictly of a gravitational nature.\\

Analyzing the roots of $f(r)$, it is identified that this black hole in MOG exhibits two horizons located at
\begin{equation}
    r_\pm = G_N M \left[ 1 + \alpha \pm \sqrt{1+\alpha} \right]. \label{eq:horizons}
\end{equation}

As a natural consistency check, it is evident that by vanishing the MOG deformation parameter (taking the limit $\alpha \to 0$), the metric reduces exactly to the Schwarzschild solution, characterized by a single horizon located at the radius $r_+ = r_s = 2G_N M$.

\section{Master equations for the different fields}

\label{sec:perturbations}

\subsection{Perturbations of a test scalar field}

\label{sec:scalar_perturbations_mog}

To study the linear evolution of a test scalar field in the spacetime of a static black hole, we start from the generalized Klein-Gordon equation \cite{berti2009quasinormal}. Considering a massive scalar field $\Phi$ of mass $\mu$, the dynamics in the absence of self-interactions is dictated by
\begin{equation}
    \left( \square - \mu^2 \right) \Phi = 0,
\end{equation}
which can be written in covariant form as
\begin{equation}
 \frac{1}{\sqrt{-g}} \partial_\gamma \left( \sqrt{-g} g^{\gamma\nu} \partial_\nu \Phi \right) - \mu^2 \Phi = 0,
    \label{eq:KleinGordon}
\end{equation}
where $g$ represents the determinant of the background metric tensor $g_{\gamma\nu}$.\\

Taking advantage of the spacetime isometries, it is possible to separate the variables of the wave equation. We decompose the scalar field $\Phi$ into harmonic modes of frequency $\omega$ and angular multipoles using the spherical harmonics $Y_{\ell m}(\theta, \varphi)$,
\begin{equation}
    \Phi(t,r,\theta,\varphi) = \sum_{\ell=0}^{\infty} \sum_{m=-\ell}^{\ell} \frac{\Psi_{\ell}(r)}{r} Y_{\ell m}(\theta, \varphi) e^{-i\omega t}.
    \label{eq:AnsatzEscalar}
\end{equation}

Substituting Eq. \eqref{eq:AnsatzEscalar} into Eq. \eqref{eq:KleinGordon}, the temporal and angular dependence is completely factorized. To bring the resulting radial equation into a standard form, we introduce the generalized ``tortoise'' coordinate, $r_*$, defined by $dr_* = dr / f(r)$. This transformation maps the exterior domain of the black hole, bounded by the outer event horizon $r_+ = G_NM\left[1+\alpha + \sqrt{1+\alpha}\right]$,  to the entire real line $r_* \in (-\infty, \infty)$. Under this transformation, the equation of motion for the radial component takes the form of a one-dimensional Schrödinger equation,
\begin{equation}
    \frac{d^2\Psi_{\ell}}{dr_*^2} + \left[ \omega^2 - V_{\ell}(r) \right]\Psi_{\ell} = 0.
    \label{eq:MasterEquation}
\end{equation}
The interaction of the scalar field with the  curvature is governed by the effective potential $V_\ell(r)$, which is obtained explicitly by evaluating the MOG metric,
\begin{equation}
    V_{\ell}(r) = f(r) \left[ \frac{\ell(\ell+1)}{r^2} + \frac{f'(r)}{r} + \mu^2 \right].
    \label{eq:EffectivePotential}
\end{equation}

The Master Equation \eqref{eq:MasterEquation} and its effective potential \eqref{eq:EffectivePotential} capture the full dynamics of the scalar perturbations. As observed in the expression for the potential, the modified gravity parameter $\alpha$ not only increases the size of the event horizon, but also modifies the curvature coupling through the $f'(r)/r$ term, altering the quasinormal mode spectrum.

\subsection{Perturbations of a test electromagnetic field}

\label{sec:em_perturbations_mog}

To extend our analysis to perturbations of  a vector field $A_\mu$, with spin $s=1$, we consider a test electromagnetic field propagating in the background of the Schwarzschild-MOG black hole. We assume that this field does not interact non-linearly with the intrinsic massive vector field of the MOG theory. The evolution of the field is governed by the covariant Maxwell equations in vacuum,
\begin{equation}
    \nabla_\mu F^{\mu\nu} = 0,
\end{equation}
which, written in covariant form, are
\begin{equation}
\frac{1}{\sqrt{-g}} \partial_\mu \left( \sqrt{-g} F^{\mu\nu} \right) = 0,
    \label{eq:Maxwell}
\end{equation}
where the electromagnetic field tensor is $F_{\mu\nu} = \partial_\mu A_\nu - \partial_\nu A_\mu$. \\

Because the field is a vector field, the separation of variables requires decomposing the potential $A_\mu$ using vector spherical harmonics \cite{nollert1999quasinormal, maggiore2018gravitational}. This decomposition classifies the perturbations into two independent sectors according to their parity transformation under spatial inversion: the axial sector (odd parity, $(-1)^{\ell+1}$) and the polar sector (even parity, $(-1)^\ell$) \cite{chandrasekhar1985mathematical, ruffini1972electromagnetic}. By substituting the four-potential decomposition into Eq. \eqref{eq:Maxwell}, the conformal nature of the electromagnetic field in four dimensions induces an exact cancellation of the spherical volume factors. Defining an auxiliary radial function $\Psi_\ell(r)$ and assuming a harmonic temporal dependence $e^{-i\omega t}$, the dynamics reduces once again to a second-order differential equation. \\

Using the same generalized tortoise coordinate, $dr_* = dr / f(r)$, defined for the MOG spacetime, the radial equation transforms into the master Schrödinger equation,
\begin{equation}
    \frac{d^2\Psi_{\ell}}{dr_*^2} + \left[ \omega^2 - V_{\ell}(r) \right]\Psi_{\ell} = 0.
    \label{eq:MasterEquationEM}
\end{equation}

The effective potential for the test electromagnetic field ($s=1$) is given by
\begin{equation}
    V_\ell^{\mathrm{EM}}(r) = f(r) \frac{\ell(\ell+1)}{r^2} .
    \label{eq:EffectivePotentialEM}
\end{equation}

\subsection{Gravitational perturbations}

The analysis of gravitational perturbations in the Schwarzschild-MOG spacetime proceeds under the same canonical formalism established for General Relativity, but with the additional complexity of simultaneously perturbing the metric tensor and the intrinsic massive vector field of the theory: \\

\begin{equation}
g_{ab} = g^{(0)}_{ab} + h_{ab}, \quad |h_{\mu\nu}| \ll 1,
\label{eq:mog_metric_perturbated}
\end{equation}

\begin{equation}
\phi_{a} = \phi^{(0)}_{a} + \delta \phi_{a}, \quad |\delta \phi_{a}| \ll 1,
\label{eq: mog_vec_field_pert}
\end{equation}

Taking advantage of the spherical symmetry of the background, the metric perturbation $h_{\mu\nu}$ and the field perturbation $\delta \phi_{a}$ are decomposed using tensor \cite{nagar2005gauge, thorne1980multipole, zerilli1970effective} and vector spherical harmonics, respectively. This decomposition once again classifies the degrees of freedom into two independent sectors according to their parity: axial perturbations (odd parity) and polar perturbations (even parity) \cite{zerilli1974perturbation, zerilli1970effective, regge1957stability, nagar2005gauge, mathews1962gravitational}. To simplify the complex system of coupled differential equations, the Regge \textit{gauge} \cite{kokkotas1999quasi, nollert1999quasinormal, regge1957stability} is traditionally adopted, which fixes a reference frame that systematically eliminates the non-physical spacetime degrees of freedom.\\

A distinctive feature of MOG gravity is that the presence of the background vector field induces a strong coupling between the gravitational fluctuations and the vector fluctuations. This scenario is mathematically analogous to the Einstein-Maxwell formalism for a charged black hole. Consequently, the coupled perturbation system in MOG turns out to be mapped from the perturbation problem in the Reissner-Nordström spacetime, which allows diagonalizing the equations and establishing a direct mapping for the master equations, as detailed below. The linear analysis of coupled perturbations (gravitational and electromagnetic) in a spherically symmetric and charged spacetime was rigorously formalized by Zerilli \cite{zerilli1974perturbation} and Moncrief \cite{moncrief1974odd, moncrief1975gauge}, and later unified in the canonical work of Chandrasekhar \cite{chandrasekhar1985mathematical}. In this theoretical framework, strictly applicable to the Reissner-Nordström (RN) metric, the perturbation equations decouple into two independent master equations governing the even and odd parities. \\

The master differential equation for perturbations in the Reissner-Nordström spacetime, in Chandrasekhar's notation \cite{chandrasekhar1985mathematical}, takes the form

\begin{equation}
    \left( \frac{d^2}{dr_*^2} + \sigma^2 \right) Z_i^{(\pm)} = V_i^{(\pm)}(r)  Z_i^{(\pm)} , \quad (i=1,2),
    \label{eq:master_RN}
\end{equation}
where
\begin{align}
V_i^{(-)}(r) &= \frac{\Delta}{r^5} \left[ (\mu^2 +2) r - q_j   \left( 1 + \frac{q_i}{\mu^2 r}\right) \right]
\label{eq:pot_RN_impar}
\end{align}

\noindent for $i, j \in \{1, 2\}$ with $i \neq j$, $q_1 + q_2 = 6M_{RN}$, $-q_1q_2 = 4Q^2_{RN} \mu^2$ and

\begin{align}
\Delta  &= r^2 - 2M_{RN} r + Q^2_{RN} \label{eq:Delta_RN}\\
q_{1,2} &=   3M_{RN} \pm \sqrt{9M_{RN}^2 + 4Q_{RN}^2\mu^2} \label{eq:q_RN} \\
\mu^2 &= \ell(\ell+1) - 2 = (\ell-1)(\ell+2) \\
\sigma^2 &= \omega^2.
\end{align}

Here, $M_{RN}$ and $Q_{RN}$ represent the mass and the electric charge of the Reissner-Nordström black hole, respectively, $\mu^2$ is Chandrasekhar's angular parameter (not to be confused with the scalar-field mass of Sec.~\ref{sec:scalar_perturbations_mog}), and the radial function is \cite{chandrasekhar1985mathematical}

\begin{equation}
    f(r) = \frac{\Delta}{r^2}=1 - \frac{2M_{RN}}{r} + \frac{Q_{RN}^2}{r^2},
    \label{eq:f_RN}
\end{equation}
where $r_*$ denotes the tortoise coordinate defined by $dr_* = (r^2/\Delta_{RN})dr$, and the subscript $i$ classifies the electromagnetic ($i=1$) and gravitational ($i=2$) sectors. The coupling constants $q_j$, responsible for the spectral mixing induced by the electro-gravitational interaction of the background, exhibit a strict and non-linear dependence on the geometric parameters of the spacetime. \\

The exact, static and spherically symmetric solution in MOG, whose metric is determined by the radial function (\ref{eq:metric_mog}), allows identifying, by comparison with the RN radial function (\ref{eq:f_RN}), the effective geometric parameters of the MOG black hole,
\begin{eqnarray}
    M_{RN} &\rightarrow& G M = G_N(1+\alpha)M, \label{eq:G_newtonian} \\
    Q_{RN}^2 &\rightarrow&  \alpha (1+\alpha) G_N^2M^2. \label{eq:map_charge}
    %G_{N} &\rightarrow& G_{MOG}, \label{eq:G_newtonian} \\
    %Q_{RN}^2 &\rightarrow&  \alpha  G_NM^2, \label{eq:map_charge}
\end{eqnarray}

By substitution of these parameters into equations (\ref{eq:Delta_RN}) and (\ref{eq:q_RN}), we obtain
\begin{align}
\Delta  &= r^2 - 2(1+\alpha)G_N M r + \alpha(1+\alpha)G_N^2M^2 \label{eq:Delta_MOG} \\
q_{1,2} &=   G_N M \left[ 3(1+\alpha) \pm \sqrt{9 (1+\alpha)^2 + 4 \alpha(1+\alpha) \mu^2} \right].
\end{align}

This time, we have that
\begin{align}
    q_1 + q_2 &= 6(1+\alpha)G_N M\\
    q_1 q_2 &= - 4\alpha(1+\alpha) G_N^2 M^2 \mu^2.
\end{align}

The master equation, that governs the mixing of modes in modified gravity, becomes
\begin{equation}
    \left( \frac{d^2}{dr_*^2} + \sigma^2 \right) Z_i^{(\pm)} = V_i^{(\pm)}(r)  Z_i^{(\pm)} , \quad (i=1,2),
    \label{eq:master_mog_corrected}
\end{equation}
where
\begin{align}
V_i^{(-)}(r) &= \frac{\Delta}{r^5} \left[ A r - q_j +   \frac{4 \alpha(1+\alpha)G_N^2 M^2}{r}\right]
\label{eq:pot_MOG}
\end{align}
for $i, j \in \{1, 2\}$ with $i \neq j$ and, using the notation of Manfredi \cite{manfredi2018quasinormal}, $\mu^2 + 2 = \ell(\ell+1) = A$. \\

In recent literature addressing quasinormal modes in this modified gravity \cite{manfredi2018quasinormal}, a methodological inconsistency was incurred by directly inserting the vector field charge $Q_g^2$ into Eqs.~\eqref{eq:pot_RN_impar} and \eqref{eq:q_RN}, simultaneously omitting the gravitational amplification factor $(1+\alpha)$. Such a substitution decouples the perturbation equations from the actual background geometry, producing a formalism that evaluates perturbations on a spacetime  that does not coincide with Eq.~\eqref{eq:metric_mog}. \\

To deduce the physically consistent quasinormal mode spectrum of the theory, it is imperative to insert the parametric mapping defined by Eqs.~\eqref{eq:G_newtonian} and \eqref{eq:map_charge} into the Moncrief-Chandrasekhar coupling formalism. Following the numerical convention adopted in previous literature to facilitate computational integration and the subsequent comparison of spectra \cite{manfredi2018quasinormal}, we use geometrized units, $G_N = c = 1$, and rescale the radial coordinate by setting the bare mass parameter as $2M = 1$. Under this rescaling, the lapse function for the MOG spacetime reduces to
\begin{equation}
    f(r) = 1 - \frac{1+\alpha}{r} + \frac{\alpha(1+\alpha)}{4r^2}.
    \label{eq:f_scaled}
\end{equation}

This correction alters both the height of the potential barrier and the associated complex frequencies. All potentials, figures and numerical integrations in this work use the rescaling $2M = 1$; the quasinormal frequencies, however, are reported as $M\omega$, i.e.\ half of the value obtained in these internal units.

\section{Methodology}

\label{sec:methodology}

To determine the frequency spectrum of the quasinormal modes in the Schwarzschild-MOG spacetime, a dual numerical approach has been implemented that combines traditional semi-analytical techniques with deep learning methodologies. Throughout this analysis we work exclusively with the odd-parity (axial) sector, governed by the potential  of Eq.~\eqref{eq:pot_MOG}, since it admits a comparatively simpler effective potential and is free of the additional gauge subtleties of the even-parity equations. This choice entails no loss of generality: as established by Chandrasekhar \cite{chandrasekhar1985mathematical} for the Reissner-Nordström geometry, the axial and polar potentials are superpartners related by a Darboux transformation, so that both parity sectors share an identical quasinormal mode spectrum. By virtue of this isospectrality, the frequencies computed from the odd-parity equations are exactly those that would be obtained from their even-parity counterparts. \\

The WKB (Wentzel-Kramers-Brillouin) method adapted for the calculation of quasinormal modes is a semi-analytical technique founded on the analogy of wave scattering through a potential barrier in quantum mechanics. The algorithm proceeds by analytically matching the asymptotic expansion of the wave function at the boundaries of the domain (where incoming and outgoing waves reside) with a Taylor series expansion of the wave equation around the point where the local maximum of the effective potential is located, $r_{max}$.\\

In the present analysis, the sixth-order extension formulated by Konoplya \cite{konoplya2019higher} was employed. Unlike the first-order WKB approximation (or eikonal approximation), which depends exclusively on the height of the potential $V(r_{max})$ and its second spatial derivative, the sixth-order correction systematically incorporates terms involving up to higher-order derivatives. The quantization condition that defines the frequency spectrum is expressed by the relation
\begin{equation}
    \frac{i(\omega^2 - V_0)}{\sqrt{-2 V_2}} - \sum_{i=2}^{6}\Lambda_i = n + \frac{1}{2} ,
\end{equation}
where $n=0,1,2,...$ is the overtone number and $\Lambda_i$ are the higher-order corrections. These corrections depend on the values of the potential and its derivatives of order $2i$ evaluated at the top of the barrier. For its execution, the specialized \textit{Mathematica} package developed by Konoplya \textit{et al.} \cite{konoplya2026efficient} was used. The main virtue of this computational package lies not only in the automation of the dense tensor algebra necessary to derive the analytical potential, but in the internal application of Padé approximants on the WKB series. Given that the WKB formula is strictly an asymptotic series, the Padé ordering employed in the code stabilizes the convergence of the series, mitigating truncation errors and providing precise estimates for the complex frequencies $\omega$ when $\ell > n$; for the monopole $\ell = 0$ and for overtones with $n \geq \ell$ its accuracy degrades. \\

On the other hand, and as the main methodological innovation, a Physics-Informed Neural Network  \cite{raissi2019physics} designed to directly solve the master differential equations was developed. To avoid instabilities at the boundaries, the PINN makes use of the Asymptotic Iteration Method (AIM) transformation \cite{cho2012new}, minimizing the residual of the regularized differential equation. \\

The choice of both methods lies in their profound complementarity. While the WKB approximation is perturbative and depends on the local topology of the potential peak, the PINN evaluates the differential equation globally over the entire compact domain $\xi \in [0,1]$. This not only allows us to corroborate the complex frequencies $\omega$ of the fundamental modes, but also offers a unique analytical advantage: upon convergence, the neural network approximates and allows us to extract the continuous form of the regular eigenfunction $\chi(\xi)$ evaluated in the entire exterior space of the black hole.\\

The direct resolution of the wave equation in each of the spin cases studied, in the physical domain $r \in [r_+, \infty)$ presents severe numerical challenges due to the divergent nature of the boundary conditions, corresponding to purely incoming waves at the event horizon and purely outgoing waves at spatial infinity. To enable the computation of the Physics-Informed Neural Network, it is imperative to map the problem to a compact domain and analytically extract the singular behavior of the wave function \cite{cho2012new, cornell2022using}. The first step consists of introducing a compact coordinate $\xi$, defined as
\begin{equation}
    \xi = 1 - \frac{r_+}{r},
\end{equation}
which transforms the semi-infinite domain into the closed interval $\xi \in [0, 1]$. Under this mapping, the differential operators are related by $\frac{d}{dr_*} = P(\xi)\frac{d}{d\xi}$, where the geometric factor $P(\xi)$ absorbs the metric properties of the Schwarzschild-MOG spacetime,
\begin{equation}
    P(\xi) = \frac{\xi(1-\xi)^2 [r_+ - r_-(1-\xi)]}{r_+^2}.
\end{equation}

To regularize the differential equation, an \textit{ansatz} is imposed that explicitly factorizes the asymptotic divergences induced by the physical boundaries of the black hole. We express the radial function as
\begin{equation}
    \Psi(\xi) = A(\xi)\chi(\xi),
\end{equation}
 where the prefactor $A(\xi)$ removes the divergent asymptotic behavior at the horizon ($\xi = 0$) and at infinity ($\xi = 1$). Its explicit form differs slightly between the test-field sectors ($s = 0, 1$) and the coupled sector. For the coupled sector the outgoing factor reproduces the asymptotic behavior exactly only for $\alpha = 0$: for $\alpha > 0$ the logarithmic phase $r^{-\rho(1+\alpha)}$ of the outgoing wave, with $\rho \equiv -i\omega$, is only partially factorized, so that $\chi$ retains a weak non-analytic component near $\xi = 1$ whose weight grows with $\alpha$.\\

Substituting the \textit{ansatz} into the equation of each case, the canonical form of AIM is obtained,
\begin{equation}
    \chi'' = \lambda_0(\xi)\chi' + s_0(\xi)\chi(\xi),
\end{equation}
where the parameters vary according to the master equation and are presented in detail in Appendix \ref{app:ecuaciones_aim}.\\

To solve the non-linear eigenvalue problem that defines the complex frequencies of the quasinormal modes, the network does not learn from a pre-existing dataset, but optimizes its internal parameters by minimizing the residual of the governing differential equation evaluated at a set of spatial points (collocation points). This neural network was built as a Multilayer Perceptron (MLP) using the \textit{PyTorch} framework. Given that the weights of standard neural networks operate in the domain of real numbers, the regularized complex wave function $\chi(\xi)$ was represented by dividing the output layer into two neurons corresponding to its real and imaginary parts, $\chi_\theta(\xi) = u_\theta(\xi) + i v_\theta(\xi)$. The architecture consists of an input neuron (the compact coordinate $\xi$), five densely connected hidden layers with 50 neurons each, and a two-dimensional output layer. The hyperbolic tangent was selected as the activation function due to its smoothness and continuous higher-order differentiability, a strict requirement for calculating the Laplacian operator through automatic differentiation. The network weights were initialized using the Xavier normal scheme \cite{glorot2010understanding}. The complex frequency $\omega = \omega_R - i\omega_I$, with $\omega_I > 0$, is not a static hyperparameter, but was defined as a pair of additional trainable parameters of the optimizer. To guarantee and accelerate convergence toward the correct fundamental mode, $\omega$ was initialized using the semi-analytical estimate provided by the first-order WKB method. \\

The training of the PINN is governed by the minimization of a loss function composed of two fundamental terms: $\mathcal{L} = \mathcal{L}_{ODE} + \mathcal{L}_{triv}$. The first term encodes the physics of the system by penalizing the residual of the master equation transformed by the AIM method,
\begin{equation}
    \mathcal{L}_{ODE} = \frac{1}{N} \sum_{j=1}^{N} \left|  \frac{d^2\chi_\theta}{d\xi^2} - \lambda_0(\xi_j)\frac{d\chi_\theta}{d\xi} - s_0(\xi_j)\chi_\theta  \right|^2,
\end{equation}
evaluated at $N$ collocation points $\xi_j$. For the coupled sector the residual is multiplied by the regularizing factor $\Delta^2(\xi-1)^8$ before squaring, which removes the singular behavior of the coefficients at the endpoints. % VERIFICAR: confirmar si los codigos de s=0,1 usan el mismo factor
The second term is an artificial boundary condition introduced to prevent the neural network from collapsing into the trivial solution $\chi_\theta(\xi) = 0$ (which perfectly satisfies the homogeneous differential equation).  In this way, the  normalization of the wave function at an intermediate point of the spatiotemporal domain was forced through the penalty
\begin{equation}
    \mathcal{L}_{triv} = \left( |u_\theta(\xi_{mid})|^2 + |v_\theta(\xi_{mid})|^2 - 1 \right)^2.
\end{equation}

To mitigate the Runge phenomenon near the asymptotic boundaries and improve spectral convergence, the $N=500$ collocation points were not distributed uniformly, but were sampled using Chebyshev nodes mapped to the interval $\xi \in (0, 1)$. \\

The optimization process of the neural network was divided into two complementary phases. Phase 1 employed the \texttt{Adam} optimizer (with a learning rate of $10^{-3}$) for 5000 epochs to quickly explore the parameter space and bring $\omega$ and the network weights closer to the global basin of attraction. Phase 2 consisted of a high-precision refinement using the second-order \texttt{L-BFGS} (Limited-memory Broyden-Fletcher-Goldfarb-Shanno) optimizer with \textit{Strong Wolfe} line search conditions. The complete implementation of the PINN, together with the WKB routines and the scripts used to generate the eigenfunction reconstructions, is publicly available in a dedicated repository.\footnote{\url{https://github.com/anrojasor/QNM_Schwarzschild_MOG}}

\section{Results}
\label{sec:results}
Tables~\ref{tab:qnm_escalar}--\ref{tab:qnm_electromagnetica} in Appendix~\ref{appendixB} collect the quasinormal frequencies of the test scalar field ($s=0$), the test electromagnetic field ($s=1$) and the two modes of the coupled vector-tensor sector, the gravitational-dominated mode $Z_2$ and the vector-dominated mode $Z_1$, for $\alpha \in \{0,1,4,9\}$. The WKB column lists the fundamental mode $n=0$ and the overtones $n=1,2,3$; the PINN column lists the fundamental mode, and the overtones are marked with a dash. Frequencies are written as $\omega = \omega_R - i\omega_I$ with $\omega_I > 0$ and are given as $M\omega$ in units $G_N = c = 1$. The discrepancies between both methods are measured relative to the WKB value,
\begin{align}
  \epsilon_{\mathrm{Re}} &= 100\,\frac{|\omega_R^{\mathrm{PINN}} - \omega_R^{\mathrm{WKB}}|}{\omega_R^{\mathrm{WKB}}} , \\
  \epsilon_{\mathrm{Im}} &= 100\,\frac{|\omega_I^{\mathrm{PINN}} - \omega_I^{\mathrm{WKB}}|}{\omega_I^{\mathrm{WKB}}} ,
\end{align}
and are expressed in percent. For the scalar monopole $\ell = 0$ the WKB overtones with $n \geq 2$ lose monotonicity in $n$ and are shown for completeness only.

Physically, a monotonic and universal trend is observed across all perturbation sectors (scalar, vector, and tensor): as the modified gravity deviation parameter $\alpha$ increases, both the real oscillation frequency $\omega_R$ and the decay rate $\omega_I$ decrease. \\
In the pure general relativity regime ($\alpha=0$), the black hole exhibits higher oscillation frequencies and a faster decay. Conversely, a strongly coupled MOG black hole (e.g., $\alpha=9$) ``resonates'' with a lower pitch and possesses a significantly longer damping time. This effect is a direct consequence of how the parameter $\alpha$ expands the effective spatial scale of the potential barrier and modifies the local curvature. Likewise, as is standard in perturbation theory, an increase in the multipole number $\ell$ increases the height of the centrifugal barrier, resulting in higher real frequencies.

For the test electromagnetic field the PINN reproduces the WKB frequencies with discrepancies below 0.5\,\% in the real part and at most 2.5\,\% in the imaginary part for all multipoles and couplings, and for the scalar field with $\ell \geq 1$ both remain below 2\,\%. In the coupled sector the real part is reproduced within 2.3\,\% in every case, and the agreement improves with $\ell$, reaching 0.01--0.4\,\% for the $Z_2$ mode with $\ell = 4$ along the entire range of $\alpha$. The discrepancies concentrate in the imaginary part of the coupled modes, where they grow with $\alpha$ and reach $\epsilon_{\mathrm{Im}} \approx 17\,\%$ for $Z_2$ ($\ell = 2$, $\alpha = 9$) and $\approx 27\,\%$ for $Z_1$ ($\ell = 1$, $\alpha = 9$); for $Z_1$ with $\ell = 1$ they already amount to 6--7\,\% at weak coupling. The remaining large deviation appears in the scalar monopole $\ell = 0$, where $\epsilon_{\mathrm{Re}}$ reaches 7.6\,\% and the sixth-order WKB estimate is itself least reliable. In the coupled sector the origin of the discrepancy lies on the side of the neural network: at $\alpha = 0$ the $Z_1$ mode with $\ell = 1$ obeys the same master equation as the test electromagnetic field with $\ell = 1$, and both share the same WKB value, yet the corresponding PINN estimates differ in $\epsilon_{\mathrm{Im}}$ (6.1\,\% against 2.2\,\%). A plausible contribution is the partial factorization of the outgoing phase discussed in Sec.~\ref{sec:methodology}.

\begin{figure}[htbp]
     \centering
     % Primera fila
     \begin{subfigure}[b]{0.45\textwidth}
         \centering
         \includegraphics[width=\textwidth]{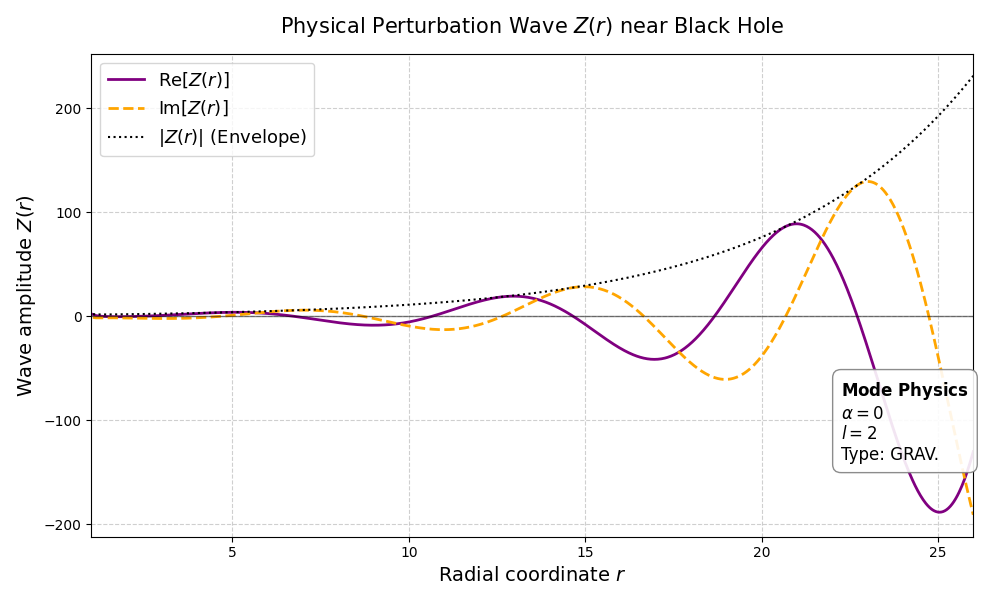}
         
     \end{subfigure}
     \hfill
     \begin{subfigure}[b]{0.45\textwidth}
         \centering
         \includegraphics[width=\textwidth]{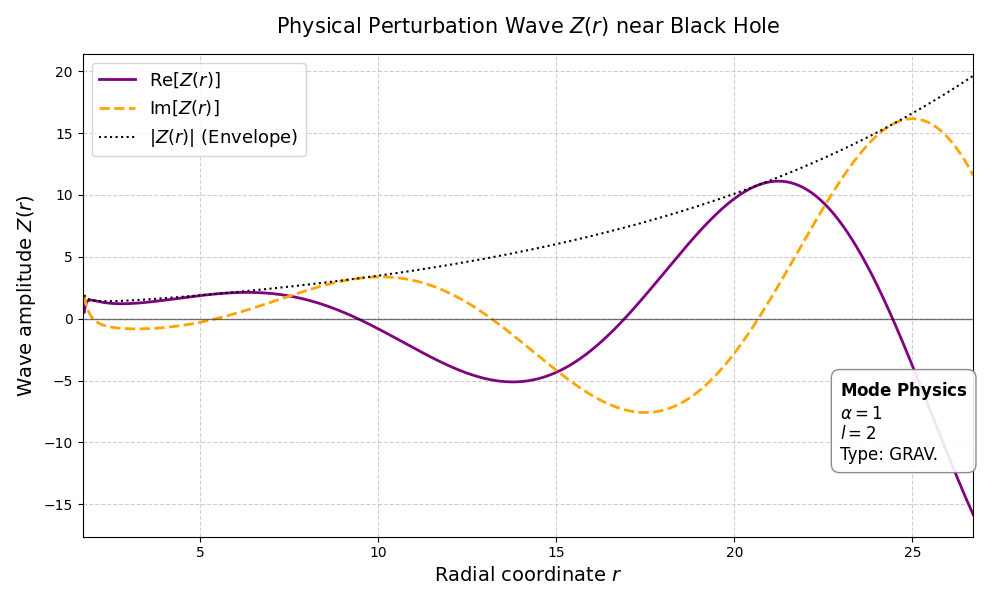}
         
     \end{subfigure}

     \vspace{10pt} % Espacio entre filas

     % Segunda fila
     \begin{subfigure}[b]{0.45\textwidth}
         \centering
         \includegraphics[width=\textwidth]{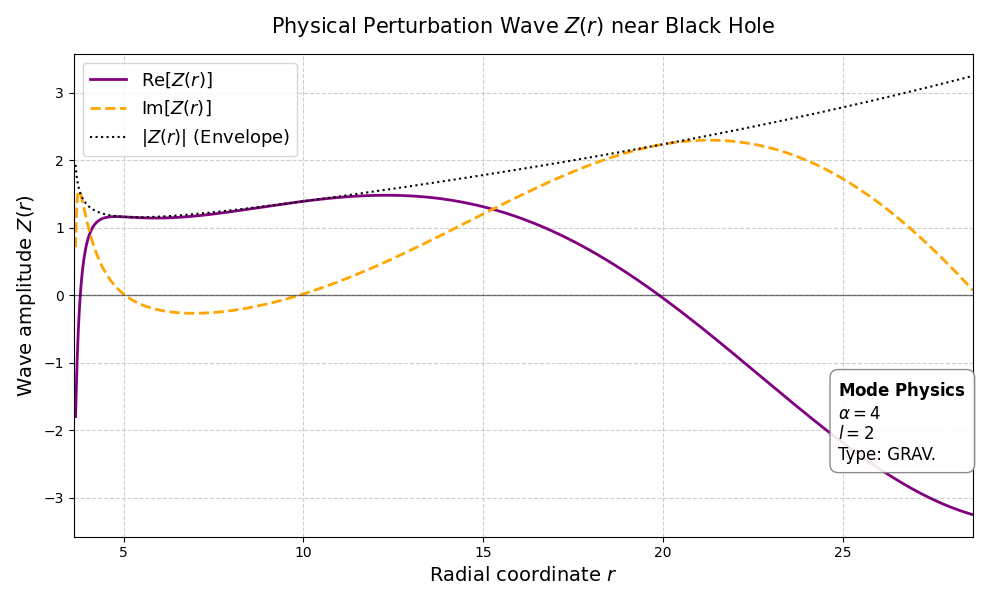}
         
     \end{subfigure}
     \hfill
     \begin{subfigure}[b]{0.45\textwidth}
         \centering
         \includegraphics[width=\textwidth]{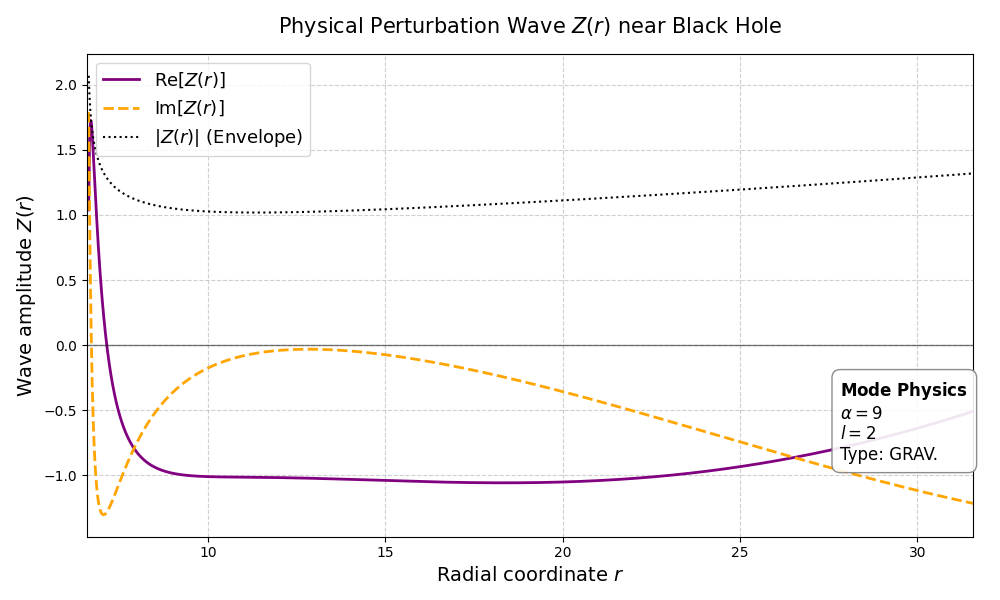}
         
     \end{subfigure}

     \caption{Eigenfunctions of the gravitational-dominated mode $Z_2$ estimated by the PINN for $n=0$, $\ell=2$ and $\alpha \in \{0, 1, 4, 9\}$.}
     \label{fig:eigenfunctions1}
\end{figure}

\begin{figure}[htbp]
     \centering
     % Primera fila
     \begin{subfigure}[b]{0.45\textwidth}
         \centering
         \includegraphics[width=\textwidth]{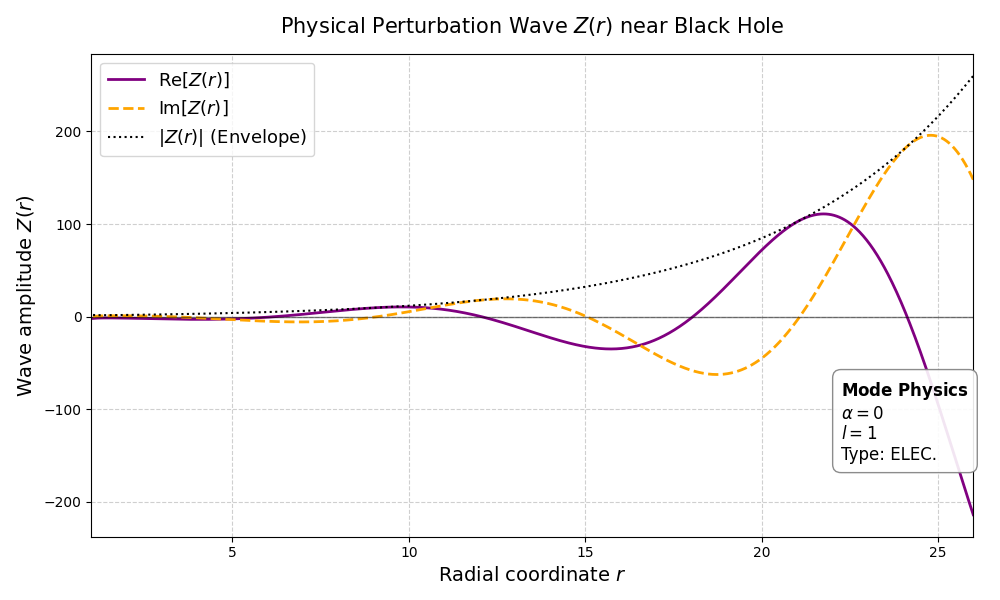}
         
     \end{subfigure}
     \hfill
     \begin{subfigure}[b]{0.45\textwidth}
         \centering
         \includegraphics[width=\textwidth]{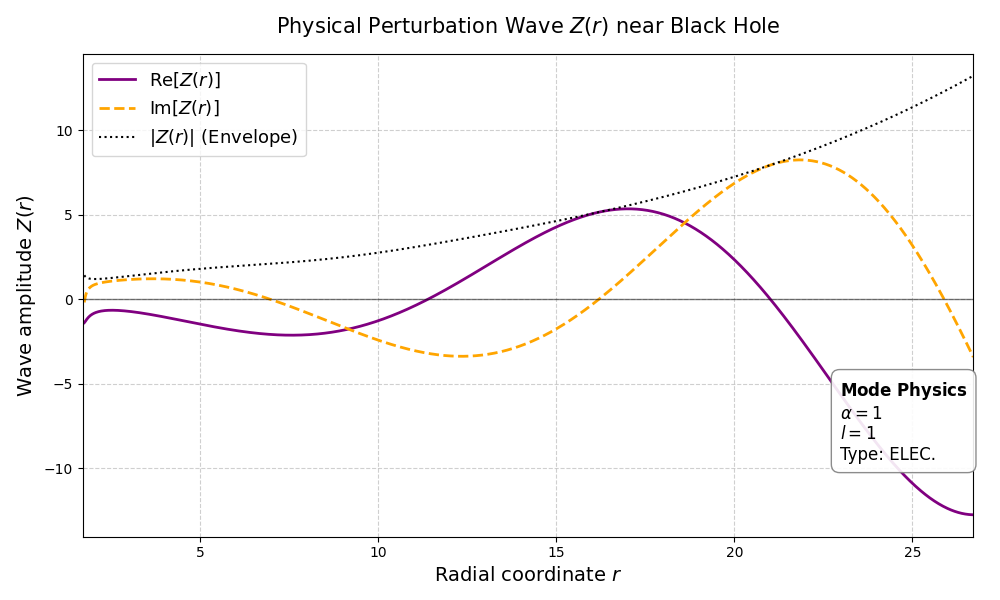}
         
     \end{subfigure}

     \vspace{10pt} % Espacio entre filas

     % Segunda fila
     \begin{subfigure}[b]{0.45\textwidth}
         \centering
         \includegraphics[width=\textwidth]{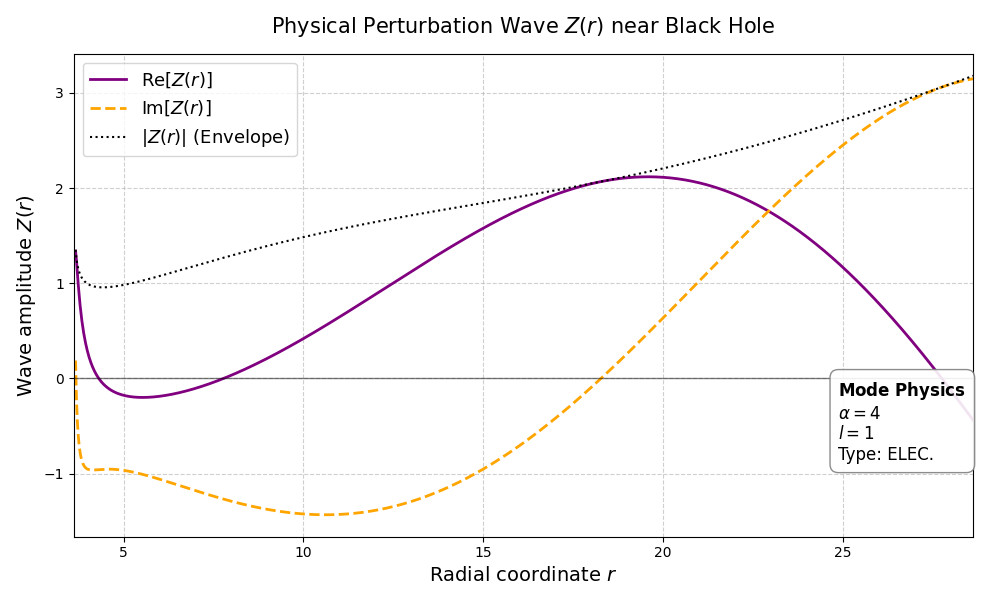}
         
     \end{subfigure}
     \hfill
     \begin{subfigure}[b]{0.45\textwidth}
         \centering
         \includegraphics[width=\textwidth]{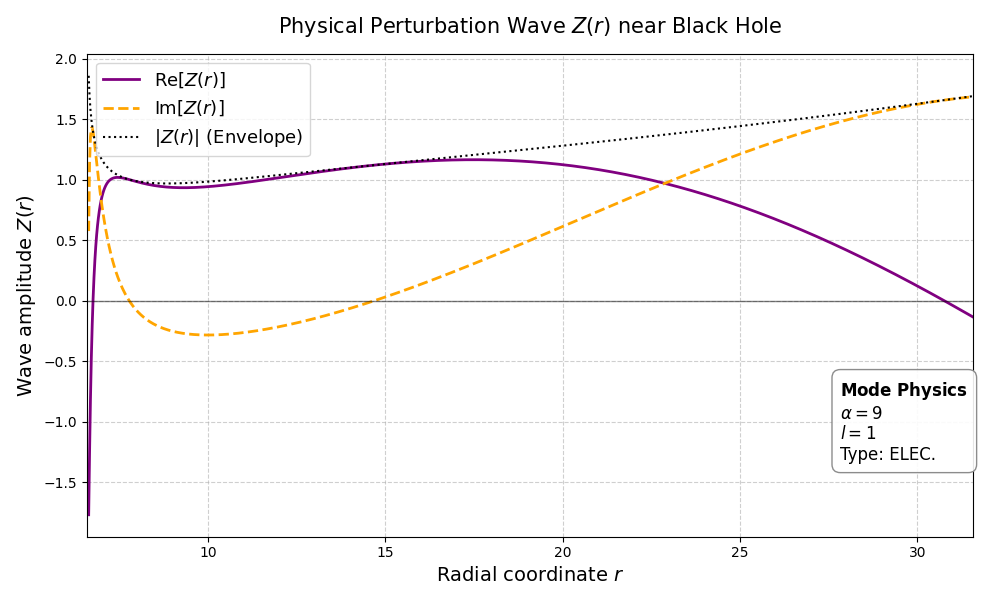}
         
     \end{subfigure}

     \caption{Eigenfunctions of the vector-dominated mode $Z_1$ of the coupled sector estimated by the PINN for $n=0$, $\ell=1$ and $\alpha \in \{0, 1, 4, 9\}$.}
     \label{fig:eigenfunctions2}
\end{figure}

An advantage of Physics-Informed Neural Networks, in contrast to purely local semi-analytical methods such as WKB, is the ability to reconstruct the global morphology of the eigenfunction over the entire exterior domain of the black hole. Figures \ref{fig:eigenfunctions1} and \ref{fig:eigenfunctions2} illustrate the spatial profiles of the real and imaginary parts of the physical reconstructed wave $Z(r) = A(\xi)\chi(\xi)$, together with its envelope $|Z(r)|$, for the fundamental gravitational-dominated mode $Z_2$ ($n=0$, $\ell=2$) and the fundamental vector-dominated mode $Z_1$ ($n=0$, $\ell=1$) under different regimes of the MOG parameter ($\alpha \in \{0, 1, 4, 9\}$). It should be noted that the network learns only the smooth, bounded part $\chi(\xi)$; the divergent asymptotic behavior characteristic of a quasinormal mode is reinjected analytically through the pre-factor $A(\xi)$, whose dominant terms are an outgoing exponential $e^{i\omega r}$ at infinity and an ingoing power-law $(r-r_+)^{-i\omega/2\kappa_+}$ at the horizon, with $\kappa_+$ the surface gravity. Since the radial coordinate is expressed in units $2M = 1$, the frequencies entering these factors are twice the tabulated values.
The global reconstruction of the regularized eigenfunctions visually reveals the impact of the MOG theory: as the deviation parameter $\alpha$ increases, the wave topology undergoes a clear spatial stretching and a reduction in its curvature, a behavior consistent with the spectral predictions of slower oscillations and more prolonged decays induced by the gravitational charge of the modified spacetime. The stretching reflects the growth of the local wavelength as the real frequency decreases, while the progressive flattening of the envelope mirrors the reduction of the decay rate, turning a steeply rising profile into a nearly flat ramp at large $\alpha$. Finally, the persistent phase lag between the real and imaginary parts of $Z(r)$ encodes the traveling-wave nature of the mode; its recovery, together with the outgoing behavior at infinity and the ingoing behavior at the horizon, confirms that the network captures the global topology of the mode rather than merely the local eigenvalue around the potential peak.

\section{Conclusions}
\label{sec:conclusions}

The present study presents a phenomenological and methodological analysis of the quasinormal mode spectrum in the Schwarzschild-MOG spacetime. At the physical level, the results consistently confirm that the presence of modified gravity (parameterized by $\alpha$) attenuates the dynamic response of the black hole, inducing lower-frequency oscillations and longer damping times, which translates into a broadening of the spatial eigenfunctions. On the analytical side, the consistent mapping between the Schwarzschild-MOG geometry and the Reissner-Nordström structure requires rescaling both the mass and the squared gravitational charge by the amplification factor $(1+\alpha)$, which modifies the spectral values reported in previous work \cite{manfredi2018quasinormal}.

However, from a computational perspective, it is necessary to adopt a prudent stance regarding the performance of the Physics-Informed Neural Network. Although the synergy of the network with the \textit{ansatz} of the AIM method proved to be conceptually successful in circumventing asymptotic divergences and reconstructing the global topology of the wave function, the obtained frequencies agree with the sixth-order WKB values with small discrepancies in most regimes, while the deviations grow in the imaginary part of the coupled modes at strong coupling and in the scalar monopole. In the coupled sector these deviations originate in the neural-network implementation, as the comparison with the test electromagnetic field at $\alpha = 0$ shows, while in the monopole they reflect the limited accuracy of WKB at $\ell = 0$. The agreement between the two methods supports the numerical implementation of both schemes; the correctness of the parametric mapping rests instead on its explicit derivation and on the limits $\alpha \to 0$. Furthermore, the current architecture showed an inherent limitation in exploring the upper spectrum: due to the spectral bias of the optimizer, the PINN preferentially converges toward the lowest energy state (the fundamental mode $n=0$), encountering severe technical difficulties in isolating and resolving the overtones ($n > 0$). Consequently, we conclude that PINNs do not yet operate as a replacement for traditional perturbative techniques, but rather as a complementary tool, in particular for the reconstruction of the eigenfunctions. A refined asymptotic factorization that includes the logarithmic phase of the outgoing wave for $\alpha > 0$ is a natural next step. While the WKB method remains the standard for calculating the spectral tower, the neural network-based approach opens a promising analytical avenue to visualize the continuous dynamics of the field and diagnose subtle spatiotemporal couplings in modified gravity theories.

 \section*{Acknowledgements} \label{sec:acknowledgements}
    This work was supported by the Universidad Nacional de Colombia, Hermes Grant Code 64813, and by the Research Incubator No.64 on Computational Astrophysics of the Observatorio Astronómico Nacional.

\nocite{jin2020unsupervised}

\bibliography{bibliography}

\newpage

% --- Inicio del Apéndice ---
\appendix

\section{Asymptotic Factorization of the Master Equations}

\label{app:ecuaciones_aim}
All expressions in this appendix use the units $G_N = c = 1$, $2M = 1$, and $\rho \equiv -i\omega$.
\subsection{Scalar perturbations}

\begin{equation}
\begin{split}
    \lambda_0 (\xi) &= - \Bigg( \frac{2\beta}{\xi} - \frac{2\gamma}{1-\xi} + \frac{2i\omega r_+}{(1-\xi)^2} \\
    &\quad + \frac{r_+(1-3\xi) - r_-(1-\xi)(1-4\xi)}{\xi(1-\xi)[r_+ - r_-(1-\xi)]} \Bigg)
\end{split}
\end{equation}
where
\begin{align}
    \beta &= -i\omega \frac{r_+^2}{r_+ - r_-}\\ 
    \gamma &= -i\omega (r_+ + r_-)\\
    k &= \sqrt{\omega^2 - \mu^2}
\end{align}

\begin{equation}
    s_0(\xi) = - \left[ \mathcal{A}(\xi) + \mathcal{B}(\xi) + \mathcal{C}(\xi) \right]
\end{equation}

\begin{equation}
\begin{split}
    \mathcal{A}(\xi) &= \left( \frac{\beta}{\xi} - \frac{\gamma}{1-\xi} + \frac{i\omega r_+}{(1-\xi)^2} \right)^2 \\
    &\quad - \frac{\beta}{\xi^2} - \frac{\gamma}{(1-\xi)^2} + \frac{2i\omega r_+}{(1-\xi)^3}
\end{split}
\end{equation}

\begin{equation}
\begin{split}
    \mathcal{B}(\xi) &= \left[ \frac{r_+(1-3\xi) - r_-(1-\xi)(1-4\xi)}{\xi(1-\xi)[r_+ - r_-(1-\xi)]} \right] \\
    &\quad \times \left( \frac{\beta}{\xi} - \frac{\gamma}{1-\xi} + \frac{i\omega r_+}{(1-\xi)^2} \right)
\end{split}
\end{equation}

\begin{equation}
    \mathcal{C}(\xi) = \frac{r_+^4}{\xi^2 (1-\xi)^4 [r_+ - r_-(1-\xi)]^2} \left[ \omega^2 - V(\xi) \right]
\end{equation}

\begin{equation}
\begin{split}
    V(\xi) &= \frac{\xi [r_+ - r_-(1-\xi)]}{r_+} \Bigg( \mu^2 + \frac{\ell(\ell+1)(1-\xi)^2}{r_+^2} \\
    &\quad + \frac{(1-\xi)^3 [r_+ - r_-(1-2\xi)]}{r_+^3} \Bigg)
\end{split}
\end{equation}

\subsection{Electromagnetic perturbations}

\begin{equation}
    s_0(\xi) = - \left[ \mathcal{A}(\xi) + \mathcal{B}(\xi) + \mathcal{C}(\xi) \right]
\end{equation}

\begin{equation}
\begin{split}
    \mathcal{A}(\xi) &= \left( \frac{\beta}{\xi} - \frac{\gamma}{1-\xi} + \frac{i\omega r_+}{(1-\xi)^2} \right)^2 \\
    &\quad - \frac{\beta}{\xi^2} - \frac{\gamma}{(1-\xi)^2} + \frac{2i\omega r_+}{(1-\xi)^3}
\end{split}
\end{equation}

\begin{equation}
\begin{split}
    \mathcal{B}(\xi) &= \left[ \frac{r_+(1-3\xi) - r_-(1-\xi)(1-4\xi)}{\xi(1-\xi)[r_+ - r_-(1-\xi)]} \right] \\
    &\quad \times \left( \frac{\beta}{\xi} - \frac{\gamma}{1-\xi} + \frac{i\omega r_+}{(1-\xi)^2} \right) % VERIFICAR contra el codigo EM
\end{split}
\end{equation}

\begin{equation}
    \mathcal{C}(\xi) = \frac{r_+^4}{\xi^2 (1-\xi)^4 [r_+ - r_-(1-\xi)]^2} \left[ \omega^2 - V(\xi) \right]
\end{equation}

\begin{equation}
\begin{split}
    V(\xi) &= f(\xi) \frac{\ell(\ell+1)(1-\xi)^2}{r_+^2} \\
    &= \frac{\xi [r_+ - r_-(1-\xi)]}{r_+} \frac{\ell(\ell+1)(1-\xi)^2}{r_+^2}
\end{split}
\end{equation}

\begin{equation}
\begin{split}
    \lambda_0 (\xi) &= - \Bigg( \frac{2\beta}{\xi} - \frac{2\gamma}{1-\xi} + \frac{2i\omega r_+}{(1-\xi)^2} \\
    &\quad + \frac{r_+(1-3\xi) - r_-(1-\xi)(1-4\xi)}{\xi(1-\xi)[r_+ - r_-(1-\xi)]} \Bigg)
\end{split}
\end{equation}

\subsection{Vector-Tensor coupled perturbations}

\begin{equation}
    \lambda_{Z_i}(\xi) = \frac{\alpha(\alpha + 1) + 4\Delta}{2\Delta(1 - \xi)} - \frac{r_+(\alpha + 2\Gamma_Z\Delta + 1)}{\Delta(\xi - 1)^2}
\end{equation}

\begin{equation}
\begin{split}
    s_{Z_i}(\xi) &= \frac{r_+^6 \left(\rho^2 + V_i^{(-)}\right)}{\Delta^2(\xi - 1)^8} - \frac{\Gamma_Z r_+^2(1 + \alpha + \Delta\Gamma_Z)}{\Delta(\xi - 1)^4} \\
    &\quad - \frac{\alpha(\alpha + 1)\Gamma_Z r_+}{2\Delta(\xi - 1)^3} - \frac{r_+\Gamma_{Z,\xi}}{(\xi - 1)^2}
\end{split}
\end{equation}

\begin{equation}
\begin{split}
    \Gamma_Z &= -\rho - \frac{1 - \xi}{r_+} + \frac{[(1 - \rho)(r_+ - r_-) - \rho r_+^2](1 - \xi)}{(r_+ - r_-)[r_+ - r_-(1 - \xi)]} \\
    &\quad + \frac{\rho r_+(1 - \xi)}{(r_+ - r_-)\xi}
\end{split}
\end{equation}

\begin{equation}
    V_i^{(-)} = \Delta \frac{(1 - \xi)^5}{r_+^5} \left[ \frac{A r_+}{1 - \xi} - q_j + \frac{4 (1+\alpha) Q_g^2(1 - \xi)}{r_+} \right]
\end{equation}

\begin{equation}
\begin{split}
    \Delta &= \frac{r_+ \xi [r_+ - r_-(1 - \xi)]}{(1 - \xi)^2} \\
    &= \frac{(1 + \alpha)\xi \left(2 + 2\sqrt{1 + \alpha} + \alpha\xi\right)}{4(\xi - 1)^2}
\end{split}
\end{equation}

\begin{equation}
    r_\pm = \frac{1}{2} \left[1 + \alpha \pm (1 + \alpha)^{1/2}\right]
\end{equation}

\begin{equation}
    Q_g = \frac{\sqrt{\alpha}}{2}
\end{equation}

\onecolumngrid

\newpage
\section{Tables of Quasinormal Frequencies in Schwarzschild-MOG} \label{appendixB}

%\clearpage % Esto enviará el siguiente título a una página nueva
%\label{app:qnm_tables} % Esta es la etiqueta
\begin{table*}[h]
\centering
\caption{Quasinormal mode spectrum for a test scalar field ($s=0$) in the Schwarzschild-MOG spacetime.}
\label{tab:qnm_escalar}
\begin{tabular}{ccccccc}
\hline
$\ell$ & $\alpha$ & $n$ & WKB & PINN & $\epsilon_{\mathrm{Re}}$ (\%) & $\epsilon_{\mathrm{Im}}$ (\%) \\
\hline
\multirow{14}{*}{0} & \multirow{2}{*}{0} & 0 & $0.11047 - 0.10082\,i$ & $0.11884 - 0.09779\,i$ & 7.58 & 3.01 \\
 &  & 1 & $0.08903 - 0.34453\,i$ & --- & --- & --- \\
 & \multirow{4}{*}{1} & 0 & $0.06116 - 0.05146\,i$ & $0.06503 - 0.04958\,i$ & 6.33 & 3.65 \\
 &  & 1 & $0.05139 - 0.17230\,i$ & --- & --- & --- \\
 &  & 2 & $0.09002 - 0.25225\,i$ & --- & --- & --- \\
 &  & 3 & $0.42777 - 0.19445\,i$ & --- & --- & --- \\
 & \multirow{4}{*}{4} & 0 & $0.02723 - 0.01904\,i$ & $0.02813 - 0.01929\,i$ & 3.31 & 1.31 \\
 &  & 1 & $0.02121 - 0.06199\,i$ & --- & --- & --- \\
 &  & 2 & $0.02494 - 0.09626\,i$ & --- & --- & --- \\
 &  & 3 & $0.09855 - 0.07439\,i$ & --- & --- & --- \\
 & \multirow{4}{*}{9} & 0 & $0.01379 - 0.00873\,i$ & $0.01446 - 0.00948\,i$ & 4.86 & 8.59 \\
 &  & 1 & $0.00860 - 0.03139\,i$ & --- & --- & --- \\
 &  & 2 & $0.00773 - 0.05812\,i$ & --- & --- & --- \\
 &  & 3 & $0.02185 - 0.07748\,i$ & --- & --- & --- \\
\hline
\multirow{16}{*}{1} & \multirow{4}{*}{0} & 0 & $0.29291 - 0.09776\,i$ & $0.29791 - 0.09751\,i$ & 1.71 & 0.26 \\
 &  & 1 & $0.26447 - 0.30652\,i$ & --- & --- & --- \\
 &  & 2 & $0.23101 - 0.54217\,i$ & --- & --- & --- \\
 &  & 3 & $0.22209 - 0.79517\,i$ & --- & --- & --- \\
 & \multirow{4}{*}{1} & 0 & $0.16177 - 0.04971\,i$ & $0.16443 - 0.04915\,i$ & 1.64 & 1.13 \\
 &  & 1 & $0.14949 - 0.15465\,i$ & --- & --- & --- \\
 &  & 2 & $0.13478 - 0.27074\,i$ & --- & --- & --- \\
 &  & 3 & $0.13070 - 0.39308\,i$ & --- & --- & --- \\
 & \multirow{4}{*}{4} & 0 & $0.07030 - 0.01948\,i$ & $0.07142 - 0.01928\,i$ & 1.59 & 1.03 \\
 &  & 1 & $0.06602 - 0.06002\,i$ & --- & --- & --- \\
 &  & 2 & $0.06014 - 0.10383\,i$ & --- & --- & --- \\
 &  & 3 & $0.05635 - 0.14962\,i$ & --- & --- & --- \\
 & \multirow{4}{*}{9} & 0 & $0.03638 - 0.00947\,i$ & $0.03677 - 0.00931\,i$ & 1.07 & 1.69 \\
 &  & 1 & $0.03415 - 0.02904\,i$ & --- & --- & --- \\
 &  & 2 & $0.03053 - 0.05009\,i$ & --- & --- & --- \\
 &  & 3 & $0.02670 - 0.07268\,i$ & --- & --- & --- \\
\hline
\multirow{16}{*}{2} & \multirow{4}{*}{0} & 0 & $0.48364 - 0.09677\,i$ & $0.48400 - 0.09587\,i$ & 0.07 & 0.93 \\
 &  & 1 & $0.46385 - 0.29563\,i$ & --- & --- & --- \\
 &  & 2 & $0.43039 - 0.50870\,i$ & --- & --- & --- \\
 &  & 3 & $0.39321 - 0.73989\,i$ & --- & --- & --- \\
 & \multirow{4}{*}{1} & 0 & $0.26691 - 0.04929\,i$ & $0.26709 - 0.04866\,i$ & 0.07 & 1.28 \\
 &  & 1 & $0.25839 - 0.15011\,i$ & --- & --- & --- \\
 &  & 2 & $0.24395 - 0.25685\,i$ & --- & --- & --- \\
 &  & 3 & $0.22786 - 0.37103\,i$ & --- & --- & --- \\
 & \multirow{4}{*}{4} & 0 & $0.11601 - 0.01936\,i$ & $0.11600 - 0.01929\,i$ & 0.01 & 0.36 \\
 &  & 1 & $0.11310 - 0.05873\,i$ & --- & --- & --- \\
 &  & 2 & $0.10802 - 0.09983\,i$ & --- & --- & --- \\
 &  & 3 & $0.10195 - 0.14311\,i$ & --- & --- & --- \\
 & \multirow{4}{*}{9} & 0 & $0.06009 - 0.00943\,i$ & $0.06007 - 0.00925\,i$ & 0.03 & 1.91 \\
 &  & 1 & $0.05865 - 0.02854\,i$ & --- & --- & --- \\
 &  & 2 & $0.05599 - 0.04835\,i$ & --- & --- & --- \\
 &  & 3 & $0.05251 - 0.06912\,i$ & --- & --- & --- \\
\hline
\end{tabular}
\end{table*}

\begin{table*}[t]
\centering
\caption{Quasinormal mode spectrum for a test electromagnetic field ($s=1$) in the Schwarzschild-MOG spacetime.}
\label{tab:qnm_electromagnetico}
\begin{tabular}{ccccccc}
\hline
$\ell$ & $\alpha$ & $n$ & WKB & PINN & $\epsilon_{\mathrm{Re}}$ (\%) & $\epsilon_{\mathrm{Im}}$ (\%) \\
\hline
\multirow{16}{*}{1} & \multirow{4}{*}{0} & 0 & $0.24819 - 0.09264\,i$ & $0.24939 - 0.09061\,i$ & 0.48 & 2.19 \\
 &  & 1 & $0.21430 - 0.29412\,i$ & --- & --- & --- \\
 &  & 2 & $0.17399 - 0.53006\,i$ & --- & --- & --- \\
 &  & 3 & $0.15900 - 0.79678\,i$ & --- & --- & --- \\
\cline{2-7}
 & \multirow{4}{*}{1} & 0 & $0.13901 - 0.04756\,i$ & $0.13910 - 0.04637\,i$ & 0.06 & 2.50 \\
 &  & 1 & $0.12452 - 0.14954\,i$ & --- & --- & --- \\
 &  & 2 & $0.10750 - 0.26538\,i$ & --- & --- & --- \\
 &  & 3 & $0.10259 - 0.39085\,i$ & --- & --- & --- \\
\cline{2-7}
 & \multirow{4}{*}{4} & 0 & $0.06132 - 0.01874\,i$ & $0.06148 - 0.01839\,i$ & 0.26 & 1.87 \\
 &  & 1 & $0.05649 - 0.05820\,i$ & --- & --- & --- \\
 &  & 2 & $0.05026 - 0.10159\,i$ & --- & --- & --- \\
 &  & 3 & $0.04696 - 0.14730\,i$ & --- & --- & --- \\
\cline{2-7}
 & \multirow{4}{*}{9} & 0 & $0.03198 - 0.00910\,i$ & $0.03201 - 0.00889\,i$ & 0.09 & 2.31 \\
 &  & 1 & $0.02958 - 0.02803\,i$ & --- & --- & --- \\
 &  & 2 & $0.02586 - 0.04858\,i$ & --- & --- & --- \\
 &  & 3 & $0.02218 - 0.07055\,i$ & --- & --- & --- \\
\hline
\multirow{16}{*}{2} & \multirow{4}{*}{0} & 0 & $0.45759 - 0.09501\,i$ & $0.45843 - 0.09397\,i$ & 0.18 & 1.09 \\
 &  & 1 & $0.43653 - 0.29073\,i$ & --- & --- & --- \\
 &  & 2 & $0.40091 - 0.50173\,i$ & --- & --- & --- \\
 &  & 3 & $0.36116 - 0.73245\,i$ & --- & --- & --- \\
\cline{2-7}
 & \multirow{4}{*}{1} & 0 & $0.25362 - 0.04852\,i$ & $0.25377 - 0.04808\,i$ & 0.06 & 0.91 \\
 &  & 1 & $0.24458 - 0.14796\,i$ & --- & --- & --- \\
 &  & 2 & $0.22927 - 0.25376\,i$ & --- & --- & --- \\
 &  & 3 & $0.21223 - 0.36760\,i$ & --- & --- & --- \\
\cline{2-7}
 & \multirow{4}{*}{4} & 0 & $0.11073 - 0.01908\,i$ & $0.11086 - 0.01877\,i$ & 0.12 & 1.62 \\
 &  & 1 & $0.10769 - 0.05794\,i$ & --- & --- & --- \\
 &  & 2 & $0.10240 - 0.09864\,i$ & --- & --- & --- \\
 &  & 3 & $0.09614 - 0.14168\,i$ & --- & --- & --- \\
\cline{2-7}
 & \multirow{4}{*}{9} & 0 & $0.05750 - 0.00929\,i$ & $0.05750 - 0.00909\,i$ & 0.00 & 2.15 \\
 &  & 1 & $0.05601 - 0.02813\,i$ & --- & --- & --- \\
 &  & 2 & $0.05328 - 0.04769\,i$ & --- & --- & --- \\
 &  & 3 & $0.04973 - 0.06825\,i$ & --- & --- & --- \\
\hline
\multirow{16}{*}{3} & \multirow{4}{*}{0} & 0 & $0.65690 - 0.09562\,i$ & $0.65704 - 0.09552\,i$ & 0.02 & 0.10 \\
 &  & 1 & $0.64174 - 0.28973\,i$ & --- & --- & --- \\
 &  & 2 & $0.61379 - 0.49206\,i$ & --- & --- & --- \\
 &  & 3 & $0.57748 - 0.70650\,i$ & --- & --- & --- \\
\cline{2-7}
 & \multirow{4}{*}{1} & 0 & $0.36321 - 0.04878\,i$ & $0.36367 - 0.04829\,i$ & 0.13 & 1.00 \\
 &  & 1 & $0.35670 - 0.14757\,i$ & --- & --- & --- \\
 &  & 2 & $0.34468 - 0.24979\,i$ & --- & --- & --- \\
 &  & 3 & $0.32906 - 0.35704\,i$ & --- & --- & --- \\
\cline{2-7}
 & \multirow{4}{*}{4} & 0 & $0.15822 - 0.01918\,i$ & $0.15821 - 0.01916\,i$ & 0.01 & 0.10 \\
 &  & 1 & $0.15603 - 0.05791\,i$ & --- & --- & --- \\
 &  & 2 & $0.15194 - 0.09763\,i$ & --- & --- & --- \\
 &  & 3 & $0.14646 - 0.13881\,i$ & --- & --- & --- \\
\cline{2-7}
 & \multirow{4}{*}{9} & 0 & $0.08208 - 0.00934\,i$ & $0.08208 - 0.00916\,i$ & 0.00 & 1.93 \\
 &  & 1 & $0.08101 - 0.02817\,i$ & --- & --- & --- \\
 &  & 2 & $0.07896 - 0.04738\,i$ & --- & --- & --- \\
 &  & 3 & $0.07611 - 0.06718\,i$ & --- & --- & --- \\
\hline
\end{tabular}
\end{table*}

\begin{table*}[t] 
\centering
\caption{Quasinormal mode spectrum of the gravitational-dominated mode $Z_2$ of the coupled vector-tensor sector in the Schwarzschild-MOG spacetime.}
\label{tab:qnm_gravitacional}
\begin{tabular}{ccccccc}
\hline
$\ell$ & $\alpha$ & $n$ & WKB & PINN & $\epsilon_{\mathrm{Re}}$ (\%) & $\epsilon_{\mathrm{Im}}$ (\%) \\
\hline
\multirow{16}{*}{2} & \multirow{4}{*}{0} & 0 & $0.37362 - 0.08889\,i$ & $0.37356 - 0.08875\,i$ & 0.02 & 0.16 \\
 &  & 1 & $0.34630 - 0.27348\,i$ & --- & --- & --- \\
 &  & 2 & $0.29852 - 0.47756\,i$ & --- & --- & --- \\
 &  & 3 & $0.24172 - 0.70960\,i$ & --- & --- & --- \\
 & \multirow{4}{*}{1} & 0 & $0.19651 - 0.04491\,i$ & $0.19625 - 0.04439\,i$ & 0.13 & 1.16 \\
 &  & 1 & $0.18373 - 0.13784\,i$ & --- & --- & --- \\
 &  & 2 & $0.16134 - 0.23960\,i$ & --- & --- & --- \\
 &  & 3 & $0.13462 - 0.35382\,i$ & --- & --- & --- \\
 & \multirow{4}{*}{4} & 0 & $0.08258 - 0.01765\,i$ & $0.08217 - 0.01685\,i$ & 0.50 & 4.53 \\
 &  & 1 & $0.07796 - 0.05384\,i$ & --- & --- & --- \\
 &  & 2 & $0.06956 - 0.09254\,i$ & --- & --- & --- \\
 &  & 3 & $0.05879 - 0.13481\,i$ & --- & --- & --- \\
 & \multirow{4}{*}{9} & 0 & $0.04216 - 0.00865\,i$ & $0.04203 - 0.00719\,i$ & 0.31 & 16.88 \\
 &  & 1 & $0.03985 - 0.02627\,i$ & --- & --- & --- \\
 &  & 2 & $0.03539 - 0.04490\,i$ & --- & --- & --- \\
 &  & 3 & $0.02914 - 0.06524\,i$ & --- & --- & --- \\
\hline
\multirow{16}{*}{3} & \multirow{4}{*}{0} & 0 & $0.59944 - 0.09270\,i$ & $0.59956 - 0.09275\,i$ & 0.02 & 0.05 \\
 &  & 1 & $0.58264 - 0.28129\,i$ & --- & --- & --- \\
 &  & 2 & $0.55159 - 0.47905\,i$ & --- & --- & --- \\
 &  & 3 & $0.51110 - 0.69047\,i$ & --- & --- & --- \\
 & \multirow{4}{*}{1} & 0 & $0.31641 - 0.04672\,i$ & $0.31645 - 0.04660\,i$ & 0.01 & 0.26 \\
 &  & 1 & $0.30870 - 0.14161\,i$ & --- & --- & --- \\
 &  & 2 & $0.29447 - 0.24061\,i$ & --- & --- & --- \\
 &  & 3 & $0.27590 - 0.34569\,i$ & --- & --- & --- \\
 & \multirow{4}{*}{4} & 0 & $0.13371 - 0.01836\,i$ & $0.13406 - 0.01781\,i$ & 0.26 & 3.00 \\
 &  & 1 & $0.13094 - 0.05552\,i$ & --- & --- & --- \\
 &  & 2 & $0.12576 - 0.09395\,i$ & --- & --- & --- \\
 &  & 3 & $0.11881 - 0.13426\,i$ & --- & --- & --- \\
 & \multirow{4}{*}{9} & 0 & $0.06848 - 0.00899\,i$ & $0.06819 - 0.00825\,i$ & 0.42 & 8.23 \\
 &  & 1 & $0.06709 - 0.02714\,i$ & --- & --- & --- \\
 &  & 2 & $0.06443 - 0.04583\,i$ & --- & --- & --- \\
 &  & 3 & $0.06072 - 0.06534\,i$ & --- & --- & --- \\
\hline
\multirow{16}{*}{4} & \multirow{4}{*}{0} & 0 & $0.80918 - 0.09416\,i$ & $0.80913 - 0.09413\,i$ & 0.01 & 0.03 \\
 &  & 1 & $0.79663 - 0.28433\,i$ & --- & --- & --- \\
 &  & 2 & $0.77270 - 0.47990\,i$ & --- & --- & --- \\
 &  & 3 & $0.73967 - 0.68390\,i$ & --- & --- & --- \\
 & \multirow{4}{*}{1} & 0 & $0.42934 - 0.04750\,i$ & $0.42939 - 0.04745\,i$ & 0.01 & 0.11 \\
 &  & 1 & $0.42365 - 0.14331\,i$ & --- & --- & --- \\
 &  & 2 & $0.41278 - 0.24153\,i$ & --- & --- & --- \\
 &  & 3 & $0.39779 - 0.34348\,i$ & --- & --- & --- \\
 & \multirow{4}{*}{4} & 0 & $0.18235 - 0.01866\,i$ & $0.18239 - 0.01861\,i$ & 0.02 & 0.27 \\
 &  & 1 & $0.18032 - 0.05623\,i$ & --- & --- & --- \\
 &  & 2 & $0.17644 - 0.09453\,i$ & --- & --- & --- \\
 &  & 3 & $0.17099 - 0.13398\,i$ & --- & --- & --- \\
 & \multirow{4}{*}{9} & 0 & $0.09360 - 0.00912\,i$ & $0.09371 - 0.00909\,i$ & 0.12 & 0.37 \\
 &  & 1 & $0.09259 - 0.02748\,i$ & --- & --- & --- \\
 &  & 2 & $0.09063 - 0.04613\,i$ & --- & --- & --- \\
 &  & 3 & $0.08782 - 0.06527\,i$ & --- & --- & --- \\
\hline
\end{tabular}
\end{table*}

\begin{table*}[t] 
\centering
\caption{Quasinormal mode spectrum of the vector-dominated mode $Z_1$ of the coupled vector-tensor sector in the Schwarzschild-MOG spacetime.}
\label{tab:qnm_electromagnetica}
\begin{tabular}{ccccccc}
\hline
$\ell$ & $\alpha$ & $n$ & WKB & PINN & $\epsilon_{\mathrm{Re}}$ (\%) & $\epsilon_{\mathrm{Im}}$ (\%) \\
\hline
\multirow{16}{*}{1} & \multirow{4}{*}{0} & 0 & $0.24819 - 0.09264\,i$ & $0.24657 - 0.08699\,i$ & 0.65 & 6.10 \\
 &  & 1 & $0.21430 - 0.29412\,i$ & --- & --- & --- \\
 &  & 2 & $0.17399 - 0.53006\,i$ & --- & --- & --- \\
 &  & 3 & $0.15900 - 0.79678\,i$ & --- & --- & --- \\
 & \multirow{4}{*}{1} & 0 & $0.15036 - 0.04890\,i$ & $0.15099 - 0.04577\,i$ & 0.42 & 6.40 \\
 &  & 1 & $0.13680 - 0.15318\,i$ & --- & --- & --- \\
 &  & 2 & $0.12089 - 0.27009\,i$ & --- & --- & --- \\
 &  & 3 & $0.11619 - 0.39458\,i$ & --- & --- & --- \\
 & \multirow{4}{*}{4} & 0 & $0.07165 - 0.01954\,i$ & $0.07141 - 0.01811\,i$ & 0.33 & 7.32 \\
 &  & 1 & $0.06766 - 0.05999\,i$ & --- & --- & --- \\
 &  & 2 & $0.06218 - 0.10312\,i$ & --- & --- & --- \\
 &  & 3 & $0.05859 - 0.14742\,i$ & --- & --- & --- \\
 & \multirow{4}{*}{9} & 0 & $0.03887 - 0.00944\,i$ & $0.03798 - 0.00686\,i$ & 2.29 & 27.33 \\
 &  & 1 & $0.03708 - 0.02869\,i$ & --- & --- & --- \\
 &  & 2 & $0.03416 - 0.04878\,i$ & --- & --- & --- \\
 &  & 3 & $0.03105 - 0.06940\,i$ & --- & --- & --- \\
\hline
\multirow{16}{*}{2} & \multirow{4}{*}{0} & 0 & $0.45759 - 0.09501\,i$ & $0.45670 - 0.09333\,i$ & 0.19 & 1.77 \\
 &  & 1 & $0.43653 - 0.29073\,i$ & --- & --- & --- \\
 &  & 2 & $0.40091 - 0.50173\,i$ & --- & --- & --- \\
 &  & 3 & $0.36116 - 0.73245\,i$ & --- & --- & --- \\
 & \multirow{4}{*}{1} & 0 & $0.26927 - 0.04941\,i$ & $0.26911 - 0.04939\,i$ & 0.06 & 0.04 \\
 &  & 1 & $0.26086 - 0.15045\,i$ & --- & --- & --- \\
 &  & 2 & $0.24659 - 0.25733\,i$ & --- & --- & --- \\
 &  & 3 & $0.23069 - 0.37156\,i$ & --- & --- & --- \\
 & \multirow{4}{*}{4} & 0 & $0.12321 - 0.01956\,i$ & $0.12313 - 0.01797\,i$ & 0.06 & 8.13 \\
 &  & 1 & $0.12062 - 0.05925\,i$ & --- & --- & --- \\
 &  & 2 & $0.11612 - 0.10050\,i$ & --- & --- & --- \\
 &  & 3 & $0.11079 - 0.14373\,i$ & --- & --- & --- \\
 & \multirow{4}{*}{9} & 0 & $0.06537 - 0.00947\,i$ & $0.06558 - 0.00931\,i$ & 0.32 & 1.69 \\
 &  & 1 & $0.06418 - 0.02863\,i$ & --- & --- & --- \\
 &  & 2 & $0.06199 - 0.04834\,i$ & --- & --- & --- \\
 &  & 3 & $0.05917 - 0.06879\,i$ & --- & --- & --- \\
\hline
\multirow{16}{*}{3} & \multirow{4}{*}{0} & 0 & $0.65690 - 0.09562\,i$ & $0.65688 - 0.09548\,i$ & 0.00 & 0.15 \\
 &  & 1 & $0.64174 - 0.28973\,i$ & --- & --- & --- \\
 &  & 2 & $0.61379 - 0.49206\,i$ & --- & --- & --- \\
 &  & 3 & $0.57748 - 0.70650\,i$ & --- & --- & --- \\
 & \multirow{4}{*}{1} & 0 & $0.38155 - 0.04951\,i$ & $0.38157 - 0.04943\,i$ & 0.01 & 0.16 \\
 &  & 1 & $0.37542 - 0.14966\,i$ & --- & --- & --- \\
 &  & 2 & $0.36413 - 0.25302\,i$ & --- & --- & --- \\
 &  & 3 & $0.34943 - 0.36105\,i$ & --- & --- & --- \\
 & \multirow{4}{*}{4} & 0 & $0.17193 - 0.01954\,i$ & $0.17201 - 0.01944\,i$ & 0.05 & 0.51 \\
 &  & 1 & $0.17001 - 0.05894\,i$ & --- & --- & --- \\
 &  & 2 & $0.16642 - 0.09920\,i$ & --- & --- & --- \\
 &  & 3 & $0.16162 - 0.14074\,i$ & --- & --- & --- \\
 & \multirow{4}{*}{9} & 0 & $0.09054 - 0.00948\,i$ & $0.09054 - 0.00911\,i$ & 0.00 & 3.90 \\
 &  & 1 & $0.08964 - 0.02855\,i$ & --- & --- & --- \\
 &  & 2 & $0.08791 - 0.04794\,i$ & --- & --- & --- \\
 &  & 3 & $0.08551 - 0.06781\,i$ & --- & --- & --- \\
\hline
\end{tabular}
\end{table*}

\end{document}